\documentclass[draft]{agujournal}
\draftfalse

\journalname{Water Resources Research}

\usepackage{amsmath,amsfonts,amssymb,bm}
\usepackage{hyperref}
\usepackage{graphicx}
\usepackage{siunitx}
\usepackage[caption=false]{subfig}
\usepackage{epstopdf}
\usepackage{booktabs}
\usepackage{cleveref} 

\DeclareSIUnit\length{L}
\DeclareSIUnit\time{T}

\begin{document}

\title{Multivariate Gaussian Process Regression for Multiscale Data Assimilation and Uncertainty Reduction}

\authors{%
  David A. Barajas-Solano\affil{1}, %
  Alexandre M. Tartakovsky\affil{1}%
}
\affiliation{1}{Pacific Northwest National Laboratory, Richland, WA}
\correspondingauthor{D. A. Barajas-Solano}{David.Barajas-Solano@pnnl.gov}

\begin{keypoints}
\item We present a parameter field estimation method based on measurements collected at different scales
\item Scales are treated as components of a multivariate Gaussian process trained from multiscale data
\item Conditioning on multiscale data can be used to reduce uncertainty in geophysical modeling
\end{keypoints}
  
\begin{abstract}
  We present a multivariate Gaussian process regression approach for parameter field reconstruction based on the field's measurements collected at two different scales, the coarse and fine scales.
  The proposed approach treats the parameter field defined at fine and coarse scales as a bivariate Gaussian process with a parameterized multiscale covariance model.
  We employ a full bivariate Mat\'{e}rn kernel as multiscale covariance model, with shape and smoothness hyperparameters that account for the coarsening relation between fine and coarse fields.
  In contrast to similar multiscale kriging approaches that assume a known coarsening relation between scales, the hyperparameters of the multiscale covariance model are estimated directly from data via pseudo-likelihood maximization.

  We illustrate the proposed approach with a predictive simulation application for saturated flow in porous media.
  Multiscale Gaussian process regression is employed to estimate two-dimensional log-saturated hydraulic conductivity distributions from synthetic multiscale measurements.
  The resulting stochastic model for coarse saturated conductivity is employed to quantify and reduce uncertainty in pressure predictions.
\end{abstract}

\section{Introduction}
\label{sec:intro}

Kriging is a widely used geostatistical tool for estimating the spatially distributed quantities from sparse measurements and for quantifying uncertainty of such estimates.
When such estimates are used as parameter fields in a physics model, uncertainty in kriging estimates can be propagated throughout the model to quantify uncertainty in predictions.
Observations of parameter fields are often gathered at different scales (often referred to as support scales).
A common example is hydraulic permeability.
Laboratory experiments performed on field samples (which are approximately \SI{0.1}{\meter} large) provide values with the support equal to the sample size.
On the other hand, indirect measurements such as cross-pumping tests have a support volume equaled to the distance between wells (\SI{10}-\SI{100}{\meter}), and correspond to local averages of the measured quantity over that support volume.
Methodologies for incorporating multiscale measurements on stochastic models for spatial parameters are important in order to systematically reduce uncertainty using all available data sources.

In~\citet{li-2009-simple}, the authors employ multivariate Gaussian process regression or cokriging, together with the properties of Gaussian processes, to formulate a~\emph{multiscale} simple kriging method for incorporating observations of the field at two scales under the assumption that coarsening is obtained by block averaging.
This multiscale simple kriging method treats the fine and coarse scale fields as covariates in bivariate cokriging, with the coarse covariance and coarse-fine cross-covariance computed analytically from the fine scale covariance in terms of the block averaging operator~\citep{vanmarcke-2010-random}.
While this approach to formulating multiscale simple kriging can be generalized to all linear coarsening operators (e.g. arithmetic moving averages, convolutions, etc.), it's not possible in general to obtain closed-form expressions for the coarse covariance and the coarse-fine cross-covariance.
Furthermore, this approach either presumes~\emph{a priori} knowledge of the structure of the coarsening operator and the coarse field support volume, which is in general application-specific and not available, or requires a sufficiently flexible parameterized linear coarsening operator with parameters to be estimated together with all other covariance hyperparameters.

As an alternative, in this work we propose to model the fine and coarse fields as components of a bivariate Gaussian process with parameterized bivariate covariance model and to estimate the shape and smoothness parameters of such model directly from multiscale data rather than deriving them from knowledge of the structure of the coarsening operator.
Multivariate covariance models are commonly used for cokriging in geostatistics.
We point the reader to~\citet{genton-2015-crosscovariance} for a comprehensive review of multivariate models.
For multiscale simple kriging, we propose employing the univariate Mat\'{e}rn kernel as the building block for multiscale covariance models as it provides sufficient flexibility for capturing the smoothness properties that arise from coarsening.
In particular, we employ the~\emph{full bivariate Mat\'{e}rn} covariance kernel proposed in~\citet{gneiting-2010-matern}.
Our approach can be extended to more than two support scales by employing the generalized construction proposed in~\citet{apanasovich-2012-valid}.
We illustrate via numerical experiments that the full bivariate Mat\'{e}rn kernel captures the different degrees of smoothness of each scale and, via multiscale simple kriging, produces accurate estimates of the field at fine and coarse scales by assimilating data collected at fine and coarse scales.  

This manuscript is structured as follows. 
In~\cref{sec:msk}, we introduce the multiscale simple kriging algorithm based on multivariate Gaussian process regression, and outline the model selection approach employed.
In~\cref{sec:coarsening}, we summarize the block averaging-based covariance model, and introduce the bivariate Mat\'{e}rn model.
Numerical experiments illustrating both inference from data and application to saturated flow modeling are presented in~\cref{sec:numerical}.
Finally, conclusions are given in~\cref{sec:conclusions}, together with possible avenues for future work.

\section{Multiscale Simple Kriging}
\label{sec:msk}

Let $\Omega \subset \mathbb{R}^d$ ($d \in [1, 3]$) be the subset of physical space of interest, and let $Y : \Omega \to \mathbb{R}$ denote a heterogeneous physical property to be estimated from discrete observations.
We assume that measurements of $Y$ are available at two different scales: (i) from soil samples (fine scale) with the measurement scale $\eta_f$ , and (ii) pumping-well tests (coarse scale) with the measurement scale $\eta_c$.
Our goal is to incorporate both sets of measurements to derive estimators of $Y$ at either support scale.

We use $y^f$ to denote $Y$ sampled at the support length-scale $\eta_f$ and $y^c$  for $Y$ sampled at the support length-scale $\eta_c$.
Next, we assume that a ``coarsening'' operator $\mathcal{R}^{cf}$ exists such that
\begin{linenomath*}
  \begin{equation}
    \label{eq:restriction}
    y^c(\mathbf{x}) \equiv [ \mathcal{R}^{cf} y^f](\mathbf{x}) \text{ for all } \mathbf{x} \in \Omega.
  \end{equation}
\end{linenomath*}
A total of $N^f$ and $N^c$ observations of $y^f$ and $y^f$ are available at locations $\{ \mathbf{x}^f_i \in \Omega \}^{N^f}_{i = 1}$ and $\{ \mathbf{x}^c_i \in \Omega \}^{N^c}_{i = 1}$, respectively.
The observation locations are arranged into the $N^f \times d$ and $N^c \times d$ matrices $X^f$ and $X^c$, respectively.
Similarly, we arrange the observation into the $N^f \times 1$ and $N^c \times 1$ column vectors $y^f_s$ and $y^c_s$, respectively.
Finally, we denote by $\mathcal{D}_s \equiv (X^f_s, y^f_s, X^c_s, y^c_s)$ the multiscale data set.

We assume that $y^f$ and $y^c$ are Gaussian processes.
For univariate simple kriging, we employ the prior processes
\begin{linenomath*}
  \begin{equation*}
    y^c(\mathbf{x}) \sim \mathcal{GP} \left ( 0, C^c(\mathbf{x}, \mathbf{x}' \mid \theta^c) \right ), \quad y^f(\mathbf{x}) \sim \mathcal{GP} \left ( 0, C^f(\mathbf{x}, \mathbf{x}' \mid \theta^c) \right ),
  \end{equation*}
\end{linenomath*}
which are updated employing the single-scale data sets $(X^c_s, y^c_s)$ and $(X^f_s, y^f_s)$, respectively~\citep{cressie-2015-geostatistics}.
Here, $C^c$ and $C^f$ denote parameterized covariance functions, with hyperparameter sets $\theta^c$ and $\theta^f$, respectively, estimated from the corresponding data sets.
For multiscale simple kriging~\citep{li-2009-simple}, we assume that the fine and coarse scale fields are components of a \emph{bivariate} Gaussian process $y(\mathbf{x}) = [ y^c(\mathbf{x}), y^f(\mathbf{x}) ]^{\top}$ with prior
\begin{linenomath*}
  \begin{equation}
  \label{eq:mks-prior}
  y(\mathbf{x}) \equiv
  \begin{bmatrix}
    y^c(\mathbf{x}) \\ y^f(\mathbf{x})
  \end{bmatrix} \sim
  \mathcal{GP} \left (%
    \begin{bmatrix}
      0 \\ 0
    \end{bmatrix},
    C(\mathbf{x}, \mathbf{x}' \mid \theta) =
    \begin{bmatrix}
      C^c  (\mathbf{x}, \mathbf{x}' \mid \theta) & C^{cf}(\mathbf{x}, \mathbf{x}' \mid \theta) \\
      C^{fc}(\mathbf{x}, \mathbf{x}' \mid \theta) & C^f  (\mathbf{x}, \mathbf{x}' \mid \theta) \\
    \end{bmatrix}
  \right )
  \end{equation}
\end{linenomath*}
to be conditioned on the multiscale data set $\mathcal{D}_s$. 
Here, $C$ denotes the parameterized multiscale covariance function, with hyperparameter set $\theta$ to be estimated from the multiscale data set.

The multiscale simple kriging-based estimation procedure is as follows: First, a ``valid'' multiscale covariance model $C(\cdot, \cdot \mid \theta)$ is selected as explained later in this section. %
Second, an estimator $\theta^{*}$ of the hyperparameters of the covariance model is computed from the multiscale data set $\mathcal{D}_s$ (model selection stage).
Finally, the estimator $\theta^{*}$ is plugged into the covariance model, and the mean and covariance of the conditional bivariate GP $y \mid \mathcal{D}_s, \theta^{*}$ are computed (plug-in conditioning stage).

The choice of multiscale covariance model and hyperparameters is not arbitrary.
First, the model must be ``valid'', i.e. the block components $C^c$, $C^f$ and $C^{cf} (\equiv (C^{fc})^{\top})$ and the hyperparameters $\theta$ must be so that $C$ is a symmetric positive definite (SPD) kernel.
Second, in a weaker sense, covariance model and hyperparameters must reflect that $y^c$ is a coarsening of $y^f$ as given by~\cref{eq:restriction}.
The approach we pursue in this work is different from the approach of~\citet{li-2009-simple} and \citet{cressie-2015-geostatistics}, where block averaging with known support is employed as the coarsening operator, from which block components can be derived such that the previous conditions are satisfied.
In contrast, in this work we propose a multiscale covariance model that can be used even when the support and/or the structure of the coarsening are unknown {\it a priori}.

For the remainder of this manuscript we employ the symbol $C(\cdot, \cdot \mid \theta)$ to denote both covariance matrices and functions, i.e., for $N^a \times d$ and $N^b \times d$ matrices $X^a$ and $X^b$ of locations in $\Omega$, $C(X^a, X^b \mid \theta)$ denotes the covariance matrix $\{ C_{ij} \equiv C(\mathbf{x}^a_i, \mathbf{x}^b_j \mid \theta) \}$.

\subsection{Conditioning on Multiscale Data}
\label{sec:msk-conditioning}

For given hyperparameters $\theta$, the posterior process conditional on the data $\mathcal{D}_s$ is $y \mid \mathcal{D}_s,\theta \sim \mathcal{GP}(\hat{\mu}, \hat{C})$,
with conditional mean and covariance given by
\begin{linenomath*}
  \begin{align}
    \label{eq:msk-cond-mean}
    \hat{\mu}(\mathbf{x} \mid \theta) &= C(\mathbf{x}, X_s \mid \theta) C^{-1}_s(\theta) y_s, \\
    \label{eq:msk-cond-covar}
    \hat{C}(\mathbf{x}, \mathbf{x}' \mid \theta) &= C(\mathbf{x}, \mathbf{x}' \mid \theta) - C(\mathbf{x}, X_s \mid \theta) C^{-1}_s (\theta) C(X_s, \mathbf{x}' \mid \theta),
  \end{align}
\end{linenomath*}
where $y^{\top}_s = [(y^c_s)^{\top}, (y^f_s)^{\top}]$ is the $N \times 1$ vector of observations, $N \equiv N^f + N^c$, and $C_s(\theta) \equiv C(X_s, X_s \mid \theta)$ denotes the covariance matrix of the observations.

From \cref{eq:msk-cond-mean} and~\cref{eq:msk-cond-covar} we can find the conditional mean and covariance for each scale.
For the coarse scale, we have $y^c \mid \mathcal{D}_s, \theta \sim \mathcal{GP}(\hat{\mu}^c, \hat{C}^c)$, with
\begin{linenomath*}
  \begin{align}
    \label{eq:msk-cond-mean-coarse}
    \hat{\mu}^c(\mathbf{x} \mid \theta) &= C^c_p(\mathbf{x}, X_s \mid \theta) C^{-1}_s(\theta) y_s, \\
    \label{eq:msk-cond-covar-coarse}
    \hat{C}^c(\mathbf{x}, \mathbf{x}' \mid \theta) &= C^c(\mathbf{x}, \mathbf{x}' \mid \theta) - C^c_p(\mathbf{x}, X_s \mid \theta) C^{-1}_s (\theta) C^c_p(X_s, \mathbf{x}' \mid \theta),\\
    \label{eq:msk-cond-p-coarse}
    C^c_p(\mathbf{x}, \mathbf{x}' \mid \theta) &=
                                                 \begin{bmatrix}
                                                   C^c(\mathbf{x}, \mathbf{x}' \mid \theta) & C^{cf}(\mathbf{x}, \mathbf{x}' \mid \theta)
                                                 \end{bmatrix}.
  \end{align}
\end{linenomath*}
Similar expressions can be found for the fine scale.

\subsection{Model Selection}
\label{sec:msk-model-selection}

For the plug-in approach employed in this work we need to compute an estimate $\theta^{*}$ of the covariance hyperparameters $\theta$ from the data set $\mathcal{D}_s$.
Two approaches are commonly used in univariate geostatistics: variogram curve fitting, and (pseudo-log)likelihood maximization.
In this work we employ the latter as experience shows that curve fitting is not adequate for estimating the smoothness of a differentiable processes \citep{stein-1999-interpolation}, which plays an important role in estimating coarse fields (see~\cref{sec:coarsening-block}).
Specifically, we take $\theta^{*} \equiv \operatorname{arg\,max}_{\theta} L(\mathcal{D}_s, \theta)$, where $L(\mathcal{D}_s, \theta)$ is a (pseudo-)log likelihood function.

Two (pseudo-)log likelihood functions are employed in this work.
The first one is the \emph{marginal likelihood} (ML), commonly used in Bayesian model selection, and defined as the likelihood of the observations given the data and the hyperparameters:
\begin{linenomath*}
  \begin{equation}
  \label{eq:ll-ml}
  L_{\mathrm{ML}}(\mathcal{D}_s, \theta) \equiv \log p(y_s \mid \mathcal{D}_s, \theta) = -\frac{1}{2} y_s C^{-1}_s(\theta) y_s - \frac{1}{2} \log | C_s(\theta) | - \frac{N}{2} \log 2 \pi.
  \end{equation}
\end{linenomath*}
The computational cost of computing the ML log likelihood is $O(N^3)$, chiefly driven by the inversion of the covariance matrix of the observations.
We note that ML estimation is equivalent to the well-known residual maximum likelihood (REML)~\citep{cressie-2015-geostatistics} estimation for the case of zero mean priors, as in~\cref{eq:mks-prior}.

The second function is the \emph{leave-one-out cross-validation} (LOO-CV) pseudo-log likelihood, which consists of sum of the log predictive probabilities of each observation given all other observations and the hyperparameters.
We denote by $\mathcal{D}_{-i} = (X_{-i}, y_{-i})$ the data set excluding the $i$th observation.
The prediction for the $i$th observation given $\mathcal{D}_{-i}$ and $\theta$ has distribution $p(y_i | \mathcal{D}_s, \theta) = \mathcal{N}(y_i \mid \hat{\mu}_{-i}, \hat{\sigma}_{-i})$, where $\hat{\mu}_{-i}$ and $\hat{\sigma}_{-i}$ are given by applying~\cref{eq:msk-cond-mean} and~\cref{eq:msk-cond-covar} to $\mathcal{D}_{-i}$, obtaining
\begin{linenomath*}
  \begin{align*}
    \hat{\mu}_{-i}(\theta) &= C(\mathbf{x}_i, X_{-i} \mid \theta) C^{-1}_{-i}(\theta) y_{-i}, \\
    \hat{\sigma}^2_{-i}(\theta) &= C(\mathbf{x}_i, \mathbf{x}_i \mid \theta) - C(\mathbf{x}_i, X_{-i} \mid \theta) C^{-1}_{-i} (\theta) C(X_{-i}, \mathbf{x}_i \mid \theta),
  \end{align*}
\end{linenomath*}
where $C_{-i}(\theta) \equiv C(X_{-i}, X_{-i} \mid \theta)$.
The LOO-CV pseudo-log likelihood is therefore given by
\begin{linenomath*}
  \begin{equation}
  \label{eq:ll-loo-cv}
  \begin{aligned}
    L_{\mathrm{LOO}}(\mathcal{D}_s, \theta) &= \sum^N_{i = 1} \log p(y_i \mid \mathcal{D}_{-i}, \theta) \\
    &= - \sum^N_{i = 1} \left \{ \frac{\left [ y_i - \hat{\mu}_{-i}(\theta) \right ]^2}{2 \hat{\sigma}^2(\theta)_{-i}} + \frac{1}{2} \log \hat{\sigma}^2(\theta)_{-i} + \frac{1}{2} \log 2 \pi \right \}.
  \end{aligned}
  \end{equation}
\end{linenomath*}
We note that computing the LOO-CV pseudo-likelihood does not require applying~\cref{eq:msk-cond-mean} and~\cref{eq:msk-cond-covar} $N$ times, and instead can be computed with $O(N^2)$ overhead from $C^{-1}_s(\theta)$.
Therefore, the computational cost of the LOO-CV pseudo-likelihood is $O(N^3)$~\citep{Rasmussen-2005-gaussian}.

\section{Multiscale Covariance Models}
\label{sec:coarsening}

In this section we present bivariate covariance models to be used in multiscale simple kriging.
The first model is the block-averaging-based model employed in~\citet{cressie-2015-geostatistics} and \citet{li-2009-simple}.
The second model is a bivariate Mat\'{e}rn covariance model.
  Both models require assuming a certain functional form for the fine-scale covariance (with hyperparameters estimated during model selection), but the block-averaging-based model requires a priori knowledge of the the structure of the coarsening, whereas the bivariate Mat\'{e}rn covariance model doesn't.

\subsection{Block Averaging-Based Model}
\label{sec:coarsening-block}

In this approach, we construct a multiscale covariance model directly from a model for the fine scale covariance $C^f$ and the relation~\cref{eq:restriction}.
We assume that $\mathcal{R}^{cf}$ is the linear operator $\mathcal{H}_{\eta_c}$ of the form
\begin{linenomath*}
  \begin{equation}
  \label{eq:block}
  y^c(\mathbf{x}) \equiv \left [ \mathcal{H}_{\eta_c} y^f \right](\mathbf{x}) \equiv \int_{\Omega} h_{\eta_c}(\mathbf{x}, \mathbf{z}) y^f(\mathbf{z}) \, \mathrm{d} \mathbf{z},
  \end{equation}
\end{linenomath*}
where $h_{\eta_c}(\mathbf{x}, \mathbf{z})$ is an averaging kernel with support $\eta_c$ satisfying $\int_{\Omega} h(\mathbf{z}, \mathbf{z}') \, \mathrm{d} \mathbf{z}' = 1$ for any $\mathbf{z} \in \Omega$.
Linearity of the coarsening operator ensures that $y^c$ is a GP.
If $h_{\eta_c}$ is known, the coarse-scale covariance and the fine-coarse-scale cross-covariance functions can be computed as
\begin{linenomath*}
  \begin{align}
    \label{eq:Cc-block}
    C^c(\mathbf{x}, \mathbf{x}' \mid \theta) &= \iint_{\Omega \times \Omega} h_{\eta_c}(\mathbf{x}, \mathbf{z}) C^f(\mathbf{z}, \mathbf{z}' \mid \theta) h_{\eta_c}(\mathbf{z}', \mathbf{x}') \, \mathrm{d} \mathbf{z} \mathrm{d} \mathbf{z}',\\
    \label{eq:Ccf-block}
    C^{cf}(\mathbf{x}, \mathbf{x}' \mid \theta) &= \int_{\Omega} h_{\eta_c}(\mathbf{x}, \mathbf{z}) C^f(\mathbf{z}, \mathbf{x}') \, \mathrm{d} \mathbf{z},
  \end{align}
\end{linenomath*}
In this model, the multiscale covariance hyperparameters are those of the fine scale covariance model.
This approach was employed in~\citet{li-2009-simple} to formulate the authors' multiscale simple kriging method, with the block averaging kernel
\begin{linenomath*}
  \begin{equation}
  \label{eq:block-kernel}
  \begin{aligned}
    h_{\eta_c}(\mathbf{z}, \mathbf{z}') &\equiv
    \begin{cases}
      | H(\mathbf{z}) |^{-1} & \text{if } \mathbf{z}' \in H(\mathbf{z}), \\
      0 & \text{otherwise}.
    \end{cases}, \\
    H(\mathbf{z}) &= \left \{ \mathbf{x} \in \Omega \mid \max \{ |x_1 - z_1|, \dots, |x_d - z_d| \} \leq \eta_c / 2 \right \} \cap \Omega,
  \end{aligned}
  \end{equation}
\end{linenomath*}
for which explicit formulae for $C^c$ and $C^{cf}$ can be found in terms of $C^f$~\citep[see][chap. 5--7]{vanmarcke-2010-random}.
It must be noted that for the block averaging kernel~\cref{eq:block-kernel}, $y^c$ is in general not isotropic, and only stationary away from $\partial \Omega$.

This approach can be employed when the support length-scale is known and when~\cref{eq:block} accurately describes the averaging relation between $y^c$ and $y^f$.
The advantage of this approach is that the number of hyperparameters is lower than for a full multiscale model (as will be seen in~\cref{sec:coarsening-matern}), thus simplifying the model selection process.

\begin{figure}[tbhp]
  \centering
  \subfloat[$\lambda_f / \eta_c = 1$]{%
    \label{fig:block-avg-vgram-1}
    \includegraphics[width=0.45\textwidth]{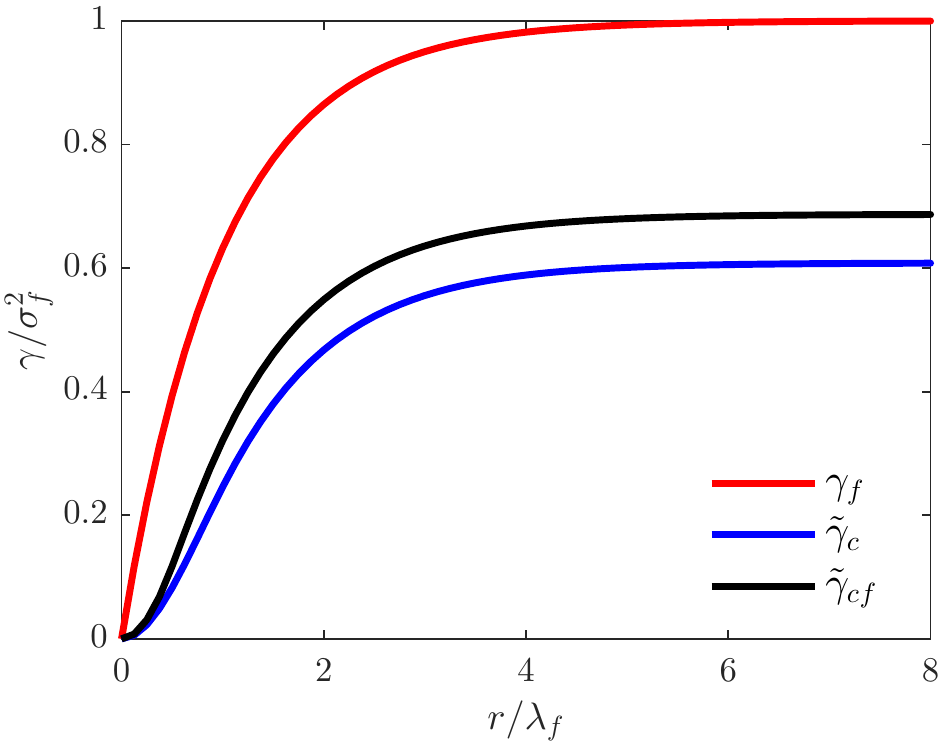}
  }
  \subfloat[$\lambda_f / \eta_c = 0.5$]{%
    \label{fig:block-avg-vgram-2}
    \includegraphics[width=0.45\textwidth]{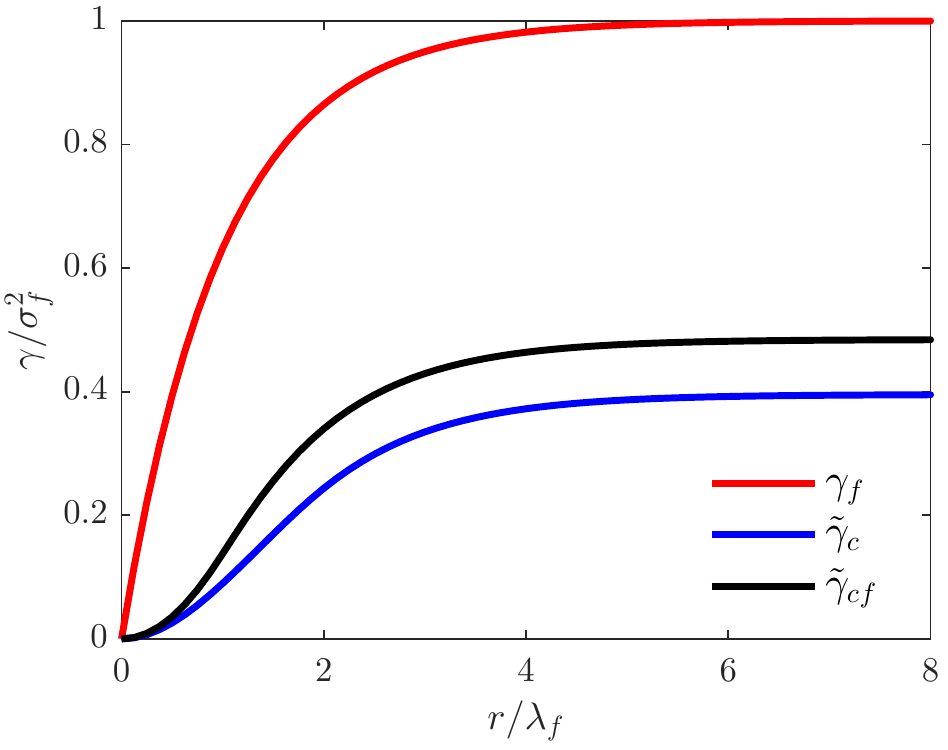}
  }
  \caption{Fine and coarse variograms, and coarse-fine cross-variogram for fine scale isotropic exponential covariance in $\mathbb{R}^2$, and two values of the coarse length-scale relative to the fine scale correlation length}
  \label{fig:block-avg-vgram}
\end{figure}

To illustrate this approach, we compute the coarse covariance and coarse-fine cross-covariance using the block averaging kernel~\cref{eq:block-kernel} for the fine scale isotropic covariance $C^f(r) = \sigma^2_f \exp (- r / \lambda_f)$ in $\mathbb{R}^2$, and for various values of $\eta_c / \lambda_f$.
Fine and coarse variograms and coarse-fine cross-variograms are presented in~\cref{fig:block-avg-vgram}.
In this case, the coarse covariance is not isotropic, therefore we present in~\cref{fig:block-avg-vgram} the pseudo-isotropic variogram $\tilde{\gamma}_c(r) \equiv C^c(\mathbf{x}, \mathbf{x}) - C^c(\mathbf{x}, \mathbf{x} + r \mathbf{e}_1)$ for any $\mathbf{x} \in \mathbb{R}^2$, where $\mathbf{e}_1$ is the unit vector along the 1st coordinate.
A similar pseudo-isotropic variogram is computed from the coarse-fine cross-covariance.

As expected, the coarse covariance exhibit longer correlation length and smaller variance than the fine scale covariance.
We also observe that the coarse correlation length increases and the coarse variance decreases with increasing $\eta_c / \lambda_f$, as noted in~\citet{tartakovsky-2017-uncertainty}.
Similar behavior is observed for the coarse-fine cross-covariance.
Therefore, applying the multiscale simple kriging estimation method described in~\cref{sec:msk} to the block-averaging model with fixed $\eta_c$ when the coarse length scale is not sufficiently well known may result in large estimation errors.
This issue can be ameliorated by including $\eta_c$ as a hyperparameter of the multiscale covariance model to be estimated during the model selection stage.
This requires assuming that the coarsening operator can be approximated by the block averaging operator with known averaging kernel.

Another effect of coarsening is in the change in smoothness from $y^f$ to $y^c$, reflected by the order of mean square differentiability of the fields.
This is seen visually in~\cref{fig:block-avg-vgram} in the behavior of $\gamma_f$ and $\tilde{\gamma}_c$ for $r \to 0$.
For the exponential kernel, we have $\gamma_f = \sigma^2_f \lambda^{-1}_f |r| + O(r^2)$ for $r \to 0$. 
According to theorem 2.7.2 in \citet{stein-1999-interpolation}, under this condition, $y^f$ is not mean square differentiable.
On the other hand, $\mathrm{d} \hat{\gamma}_c (r = 0) / \mathrm{d} r \approx 0$ (indicated by the `S'-shape of the pseudo-variogram), which indicates that $y^c$ is mean square differentiable.
This change in smoothness is due to coarsening and correspond to high correlation between $y^c$ values at locations closely together, with the radius of such high correlation region increasing with increasing $\eta_c / \lambda_f$.
Multiscale covariance models must capture this `S'-shape in order to ensure that the coarse field estimator has the appropriate degree of smoothness due to coarsening.

\subsection{Mat\'{e}rn Model}
\label{sec:coarsening-matern}

As an alternative to the block-averaging-based model presented in the previous section, we propose employing multivariate Mat\'{e}rn kernels~\citep{gneiting-2010-matern} to model block components of the multiscale covariance function.
The univariate Mat\'{e}rn kernel is given by
\begin{linenomath*}
  \begin{equation}
  \label{eq:matern}
  M(\mathbf{r} \mid \nu, \mathbf{D}) = \frac{2^{1 - \nu}}{\Gamma(\nu)} \left ( \sqrt{2 \nu} \| \mathbf{r} \|_{\mathbf{D}}  \right )^{\nu} K_{\nu} \left ( \sqrt{2 \nu} \| \mathbf{r} \|_{\mathbf{D}}  \right ),
  \end{equation}
\end{linenomath*}
where $K_{\nu}$ is the modified Bessel function of the second kind, $\nu > 0$ is a shape parameter, $\mathbf{D}$ is a $d \times d$ SPD matrix, and $\| \mathbf{r} \|_{\mathbf{D}}$ denotes the so-called Mahalanobis distance, given by
\begin{linenomath*}
  \begin{equation}
  \label{eq:mahalanobis}
  \| \mathbf{r} \|^2_{\mathbf{D}} = \mathbf{r}^{\top} \mathbf{D}^{-1} \mathbf{r}.
  \end{equation}
\end{linenomath*}
The matrix $\mathbf{D}$ allows us to introduce anisotropy to the model.
In the isotropic case, $\mathbf{D} \equiv \lambda^{-2} \mathbf{I}_d$, where $\lambda$ denotes the correlation length.
The shape parameter $\nu$ controls the smoothness of the process.
Specifically, a process with Mat\'{e}rn covariance is $m$ times mean square differentiable if and only if $\nu > m$.
The Mat\'{e}rn family includes other well known kernels as special cases:
For $\nu = 1 / 2$, the Mat\'{e}rn kernel is equal to the exponential kernel $\exp(-\| r \|_{\mathbf{D}} / \lambda)$ (not differentiable), while for $\nu \to \infty$ we recover the Gaussian or squared-exponential kernel $\exp(-\| \mathbf{r} \|^2_{\mathbf{D}} / 2 \lambda^2)$ (infinitely differentiable).

In this work we propose employing the so-called \emph{full bivariate} Mat\'{e}rn kernel as the multiscale covariance model, given by
\begin{linenomath*}
  \begin{equation}
  \label{eq:matern-full}
  C(\mathbf{r}) =
  \begin{bmatrix}
    \sigma^2_c M(\mathbf{r} \mid \nu_c, \lambda^{-2}_c \mathbf{I}_d) + \sigma^2_{nc} \bm{1}_{\| \mathbf{r} \| = 0} & \rho \sigma_c \sigma_f M(\mathbf{r} \mid \nu_{cf}, \lambda^{-2}_{cf} \mathbf{I}_d) \\
    \rho \sigma_c \sigma_f M(\mathbf{r} \mid \nu_{cf}, \lambda^{-2}_{cf} \mathbf{I}_d) & \sigma^2_f M(\mathbf{r} \mid \nu_f, \lambda^{-2}_f \mathbf{I}_d) + \sigma^2_{nf} \bm{1}_{\| \mathbf{r} \| = 0}
  \end{bmatrix}
  \end{equation}
\end{linenomath*}
where $\nu_c$, $\nu_f$ and $\nu_{cf}$ are the shape parameters for the coarse, fine and coarse-fine components, respectively; $\lambda_c$, $\lambda_f$ and $\lambda_{cf}$ are the correlation lengths of the coarse, fine, and coarse-fine components, respectively; $\sigma_c$ and $\sigma_f$ are the coarse and fine scale standard deviations, respectively, and $\rho$ is the collocated correlation coefficient.
The nugget terms $\sigma^2_{nc} \bm{1}_{\| \mathbf{r} \| = 0}$ and $\sigma^2_{nf} \bm{1}_{\| \mathbf{r} \| = 0}$ are added to the block diagonal components to capture Gaussian measurement noise.
The full bivariate model, by including separate standard deviation, shape parameters and correlation lengths for the coarse and fine scale components, provides enough flexibility to capture the effects of coarsening discussed in the previous section, namely the changes in standard deviation, correlation length, and degrees of smoothness, from the fine to coarse scale.

We proceed to establish conditions for the hyperparameters that ensure that the multiscale covariance model is valid, that is, that the full bivariate Mat\'{e}rn function~\cref{eq:matern-full} is SPD.
To simplify the full bivariate model, we consider exclusively the case $\nu_{cf} = \frac{1}{2}(\nu_c + \nu_f)$.
Numerical experiments show us that this case provides sufficient flexibility for modeling multiscale covariance functions for sufficient scale separation between $y^f$ and $y^c$, i.e. for $\eta_c / \lambda_f \gtrsim 1$.
For this case, sufficient conditions for validity are~\citet{gneiting-2010-matern}
\begin{linenomath*}
  \begin{gather}
    \label{eq:a-cond}
    a^2_{cf} \geq \frac{1}{2}(a^2_c + a^2_f), \quad a_c = \frac{\sqrt{2 \nu_c}}{\lambda_c}, \quad a_f = \frac{\sqrt{2 \nu_f}}{\lambda_f}, \quad a_{cf} = \frac{\sqrt{2 \nu_{cf}}}{\lambda_{cf}}, \text{ and}\\
    \label{eq:rho-cond}
    | \rho | \leq \frac{a_c^{\nu_c} a_f^{\nu_f}}{a^{2 \nu_{cf}}_{cf}} \frac{\Gamma ( \nu_{cf} )}{ \Gamma^{1 / 2} (\nu_c) \Gamma^{1 / 2} (\nu_f) }.
  \end{gather}
\end{linenomath*}
The number of hyperparameters to estimate during the model selection stage is 10 ($\lambda_c$, $\lambda_f$, $\lambda_{cf}$, $\nu_c$, $\nu_f$, $\sigma_c$, $\sigma_f$, $\rho$, $\sigma_{nc}$, and $\sigma_{nf}$).
In general, we expect that $\sigma_c < \sigma_f$, $\nu_c > \nu_f$, $\lambda_c > \lambda_f$, and $\rho > 0$.
These are not validity conditions and rather reflect the fact that $y^c$ is a coarsening of $y^f$ that is more smooth and has lower variance, and that $y^c$ and $y^f$ are by definition positively correlated.

The advantage of employing the full bivariate Mat\'{e}rn kernel as a multiscale covariance model is that it does not require \emph{a priori} knowledge of the coarsening relation (e.g., whether is a linear coarsening, the analytic form of $h_{\eta_c}$, the value of $\eta_c$, etc.).
Instead, the shape of the block variogram components is estimated from data, and reflected by the shape parameters $\nu_c$, $\nu_f$, and $\nu_{cf}$.
The disadvantages are that the number of free hyperparameters is larger than for the block averaging model, and that model selection requires constrained minimization of the pseudo-likelihood subject to the nonlinear constraints~\cref{eq:a-cond} and~\cref{eq:rho-cond}, as opposed to unconstrained minimization.

\section{Numerical Experiments}
\label{sec:numerical}

In this section we illustrate the proposed multiscale simple kriging approach that combines cokriging with the full bivariate Mat\'{e}rn kernel.
We employ the proposed approach to estimate two-dimensional distribution of hydraulic conductivity $K$ from synthetic multiscale hydraulic conductivity measurements.
We model $K$ at the fine scale as a stationary lognormal process, i.e., $K = K_G\exp(Y)$, where $K_G$~[\si{\length\per\time}] is the uniform geometric mean, assumed known, and $Y$, referred to as the normalized log-conductivity, is a zero-mean Gaussian process with isotropic exponential covariance.

\subsection{Model Selection}
\label{sec:numerical-selection}

We illustrate the model selection process by considering two sets of values of $\eta_f$ and $\eta_c$, corresponding to two degrees of scale separation between $y^f$ and $y^c$. 
For each test case, we obtain hyperparameter estimates using both the ML and LOO-CV approaches from multiscale observations.
We compare the estimated coarse and fine variograms against the corresponding empirical variograms computed using the \textsc{R} package \textsc{RandomFields}~\citep{randomfields,schlather-2015-analysis}.
We also compare against the true fine scale variogram and the coarse and coarse-fine pseudo-isotropic variograms.

For this experiment, the estimation domain is $\Omega = [0, 2] L \times [0, 1] L$, where $L$~[\si{\length}] is the domain length scale.
A synthetic reference fine scale log-conductivity field is simulated on a $256 \times 128$ square grid.
We take the fine support scale to be equal to the grid resolution $\Delta x$, that is, $\eta_f = \Delta x$.
The reference coarse scale log-conductivity field is then computed by block averaging with $\eta_c = m \Delta x$, where $m$ is a positive integer.
At each scale, sampling locations are randomly chosen from the set of discretization centroids.
White noise with variance $\sigma^2_{\epsilon}$ is added to the observations at both scales. All parameters and the number of measurements are listed in in~\cref{tbl:model-selection-tests}.

Constrained minimization of the negative log pseudo-likelihood is performed employing the \textsc{MATLAB} implementation of the interior point algorithm with a BFGS approximation of the Hessian.
Correlation length, shape, and standard deviation hyperparameters are log-transformed to change their support from $(0, \infty)$ to $(-\infty, \infty)$.
Similarly, the collocated correlation coefficient is transformed as $\xi = \log [ (1 + \rho) / (1 - \rho)]$ to change their support from $(-1, 1)$ to $(-\infty, \infty)$.
Initial values for the hyperparameters are chosen by inspection of the empirical variograms, and modified if the constrained minimization algorithm fails to converge.

\begin{table}[tbhp]
  \footnotesize
  \caption{Model selection tests}
  \begin{center}\footnotesize
    \begin{tabular}{*{8}{c}}
      \toprule
      & \multicolumn{4}{c}{Fine} & \multicolumn{3}{c}{Coarse}\\
      \cmidrule(lr){2-5}%
      \cmidrule(lr){6-8}%
      Test & $\sigma_f$ & $\lambda_f / L$ & \# observations $N^f$ & $\sigma_{\epsilon}$ & $m = \eta_c / \eta_f$ & \# observations $N^c$ & $\sigma_{\epsilon}$\\
      \midrule
      1  & \num{1.0} & \num{0.05} & \num{50} & \num{5e-2} & \num{8}  & \num{150} & \num{5e-2} \\
      2 & \num{1.0} & \num{0.10} & \num{40} & \num{5e-2} & \num{16} & \num{120} & \num{5e-2} \\
      \bottomrule
    \end{tabular}
  \end{center}
  \label{tbl:model-selection-tests}
\end{table}

\begin{table}[tbhp]
  \footnotesize
  \caption{%
    True and estimated full bivariate Mat\'{e}rn parameters for Tests 1 and 2 of \cref{tbl:model-selection-tests}.
    True parameters in italics ($\sigma_c$ and $\rho$) are computed from~\cref{eq:Cc-block} and \cref{eq:Ccf-block}%
  }
  \begin{center}\footnotesize
    \begin{tabular}{*{11}{l}}
      \toprule
      \multicolumn{11}{c}{Test 1}\\
      \midrule
      & $\lambda_f$ & $\sigma_f$ & $\nu_f$ & $\lambda_c$ & $\sigma_c$ & $\nu_c$ & $\lambda_{cf}$ & $\rho$ & $\sigma_{nf}$ & $\sigma_{nc}$ \\
      \midrule
      True & 0.05 & 1 & 0.5 & --- & \itshape 0.736 & --- & --- & \itshape 0.852 & 0 & 0 \\
      ML & 0.0675 & 1.04 & 0.809 & 0.0922 & 0.772 & 2.91 & 0.0846 & 0.832 & \num{9.84e-06} & \num{2.62e-07} \\
      LOO-CV & 0.0654 & 1.05 & 0.980 & 0.0877 & 0.806 & 2.99 & 0.0801 & 0.855 & \num{8.03e-06} & \num{2.92e-17} \\
      \midrule
      \multicolumn{11}{c}{Test 2}\\
      \midrule
      True & 0.1 & 1 & 0.5 & --- & \itshape 0.853 & --- & --- & \itshape 0.925 & 0 & 0 \\
      ML & 0.132 & 0.958 & 0.783 & 0.149 & 0.914 & 1.29 & 0.142 & 0.973 & \num{1.82e-06} & \num{2.99e-09} \\
      LOO-CV & 0.108 & 0.891 & 0.997 & 0.124 & 0.84 & 1.69 & 0.117 & 0.968 & \num{4.45e-07} & \num{9.86e-08} \\
      \bottomrule
    \end{tabular}
  \end{center}
  \label{tbl:model-selection-tests-hypo}
\end{table}

\begin{figure}[tbhp]
  \centering
  \subfloat[$y^f$]{%
    \includegraphics[width=0.436\textwidth]{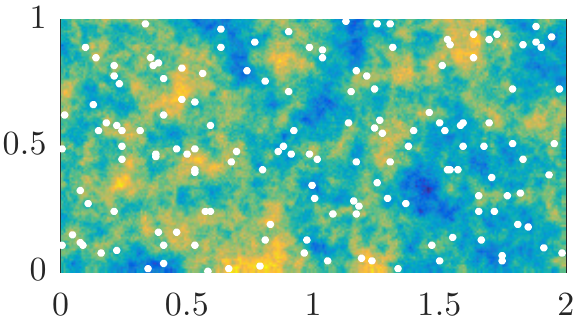}
    \label{fig:model-selection-test1-ref-yf}
  }%
  \subfloat[$y^c$]{%
    \includegraphics[width=0.544\textwidth]{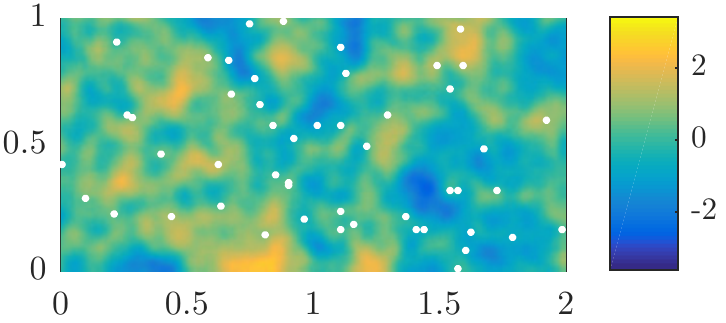}
    \label{fig:model-selection-test1-ref-yc}
  }%
  \caption{Synthetic reference fine and coarse fields, and observation locations, for Test 1}
  \label{fig:model-selection-test1-ref}
\end{figure}

\begin{figure}[tbhp]
  \centering
  \subfloat[Fine scale]{%
    \includegraphics[width=0.75\textwidth]{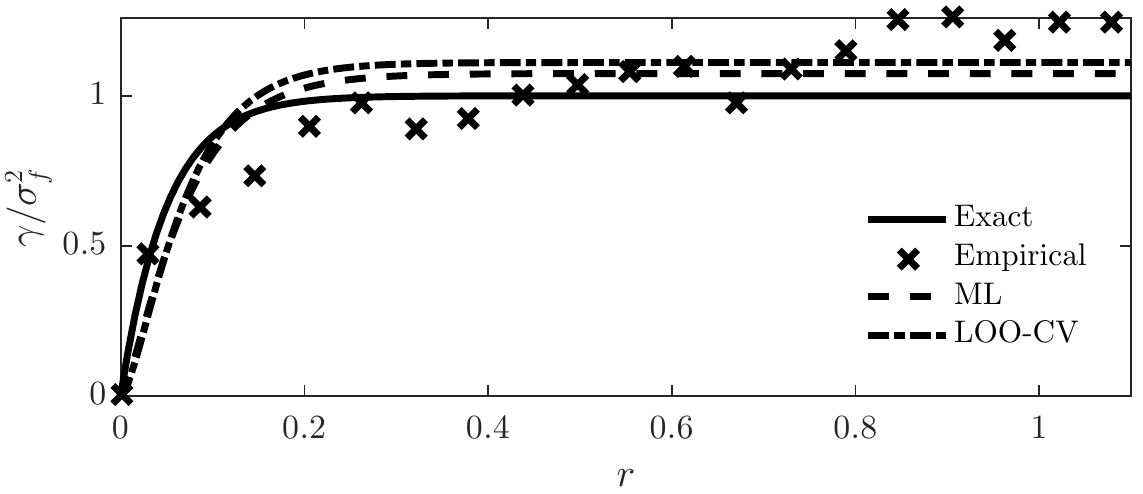}
    \label{fig:model-selection-test1-vgram-f}
  }\\
  \subfloat[Coarse scale]{%
    \includegraphics[width=0.75\textwidth]{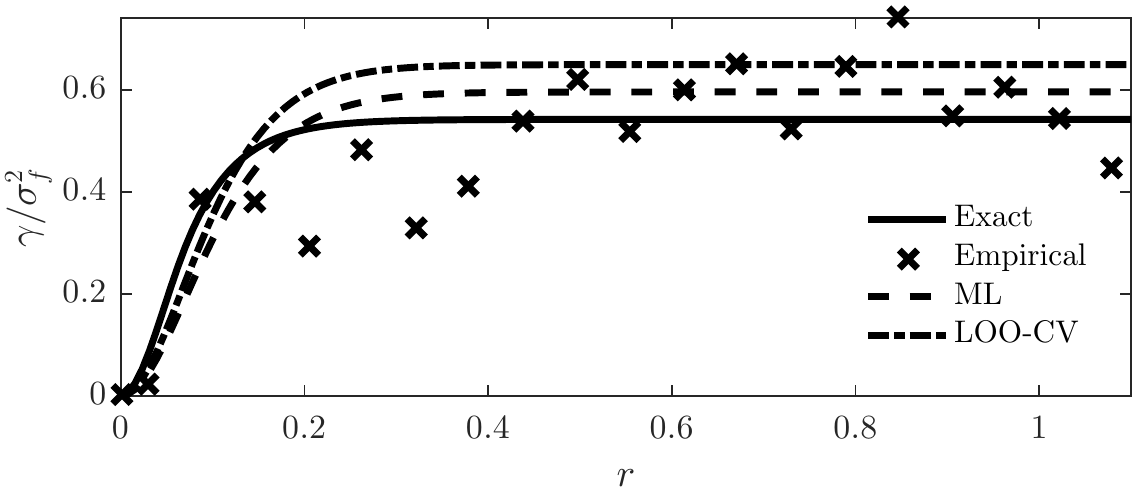}
    \label{fig:model-selection-test1-vgram-c}
  }\\
  \subfloat[Coarse-fine]{%
    \includegraphics[width=0.75\textwidth]{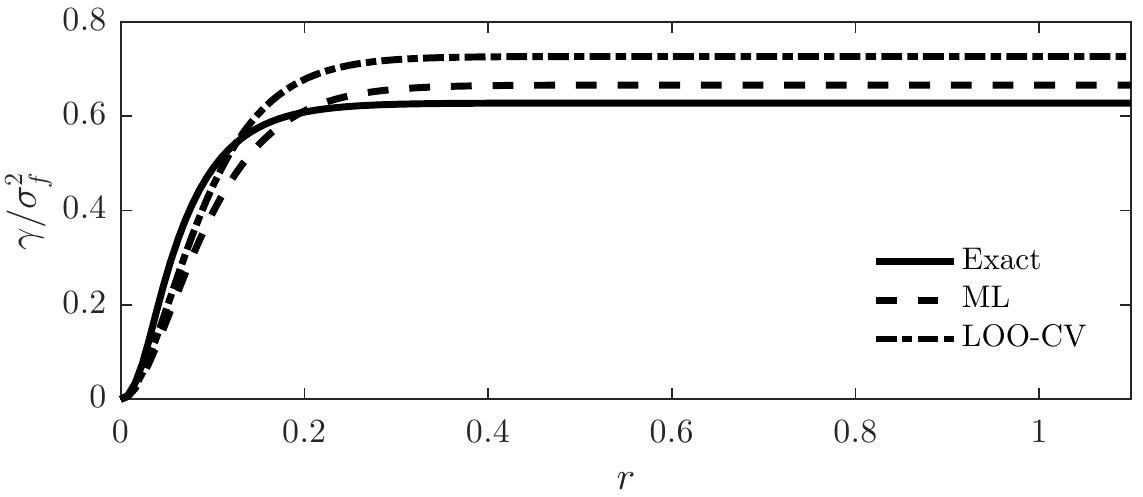}
    \label{fig:model-selection-test1-vgram-cf}
  }%
  \caption{Fine scale and coarse scale variograms, and coarse-fine cross-variograms for Test 1.
    True and estimated variograms indicated in continuous and dashed lines, respectively.
    Empirical variograms indicated with crosses.
  }
  \label{fig:model-selection-test1-vgram}
\end{figure}

\cref{tbl:model-selection-tests-hypo} presents the estimated full bivariate Mat\'{e}rn parameters for both test cases.
For Test 1, \cref{fig:model-selection-test1-ref} shows both the reference fine and coarse scale fields and the observation locations.
Similarly, \cref{fig:model-selection-test1-vgram} presents the estimated, true, and empirical variograms.
For this test we have significant scale separation as shown by the $46\%$ variance decrease from $y^f$ to $y^c$. 
From~\cref{tbl:model-selection-tests-hypo} and~\cref{fig:model-selection-test1-vgram}, we see that both ML and LOO-CV provide accurate estimates of 
the fine scale and coarse scale standard deviations and the collocated correlation coefficient, with LOO-CV overestimating the standard deviations.
Both ML and LOO-CV overestimate $\lambda_f$ by approximately $30\%$, which indicates that more fine-scale observations are necessary to more accurately estimate the fine-scale correlation length.
More importantly, both ML and LOO-CV correctly identify the difference in smoothness between scales.
Even though ML and LOO-CV overestimate the fine scale shape parameter, both correctly estimate that the fine scale field is not mean-square differentiable.
Furthermore, both methods estimate coarse correlation length and shape parameter larger than their fine scale counterparts, aligned with the observations in~\cref{sec:coarsening}.

\begin{figure}[tbhp]
  \centering
  \subfloat[Conditional mean $\hat{\mu}^f$]{%
    \includegraphics[width=0.436\textwidth]{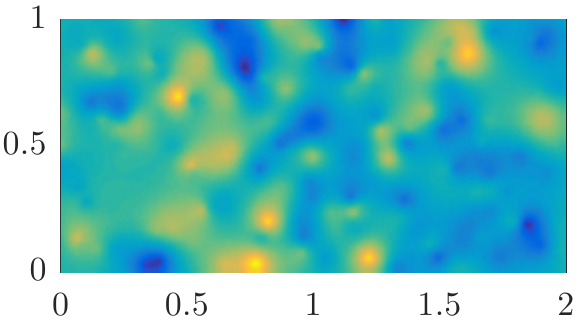}
    \label{fig:model-selection-test1-pred-mean-fine}
  }%
  \subfloat[Conditional mean $\hat{\mu}^c$]{%
    \includegraphics[width=0.544\textwidth]{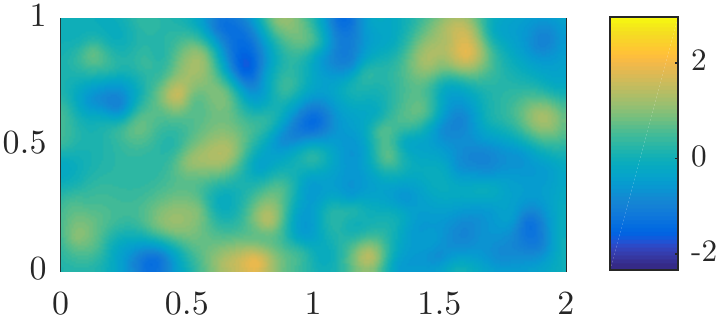}
    \label{fig:model-selection-test1-pred-mean-coarse}
  }\\
  \subfloat[Conditional variance $(\hat{\sigma}^f)^2$]{%
    \includegraphics[width=0.433\textwidth]{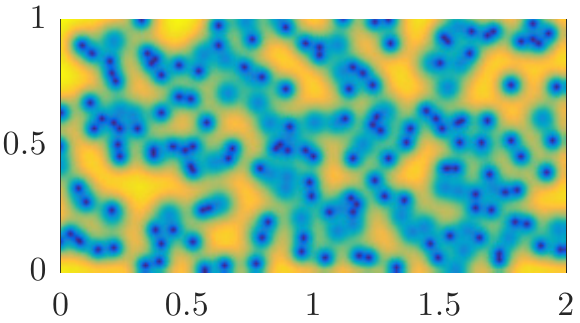}
    \label{fig:model-selection-test1-pred-var-fine}
  }%
  \subfloat[Conditional variance $(\hat{\sigma}^c)^2$]{%
    \includegraphics[width=0.547\textwidth]{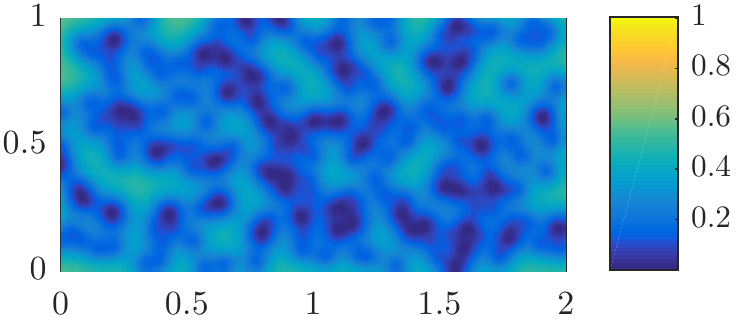}
    \label{fig:model-selection-test1-pred-var-coarse}
  }%
  \caption{Mean and variance of the fine and coarse scale fields, conditioned on multiscale data, for Test 1.}
  \label{fig:model-selection-test1-pred-meanvar}
\end{figure}

We compute the mean and variance for the fine and coarse field conditioned on multiscale data by employing the Nystr\"{o}m method presented in~\cref{sec:nystrom}.
The results are shown in~\cref{fig:model-selection-test1-pred-meanvar}.
It can be seen in~\cref{fig:model-selection-test1-pred-var-fine} and~\cref{fig:model-selection-test1-pred-var-coarse} how multiscale measurements inform the conditioning for both scales.
The measurements at a given scale reduce the conditional variance to the level of noise, while the measurements at the other scale reduce variance to a lesser but still noticeable degree. 

\begin{figure}[tbhp]
  \centering
  \subfloat[$y^f$]{%
    \includegraphics[width=0.436\textwidth]{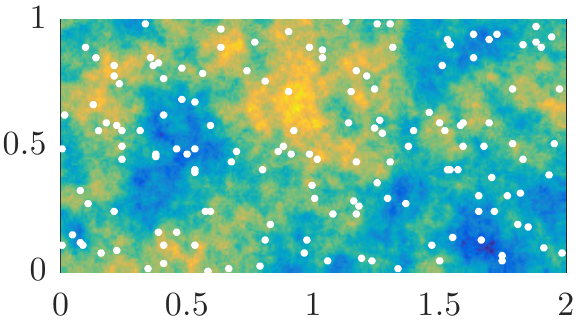}
    \label{fig:model-selection-test2-ref-yf}
  }%
  \subfloat[$y^c$]{%
    \includegraphics[width=0.543\textwidth]{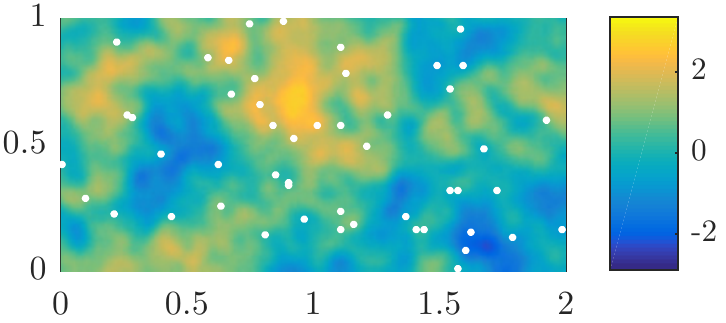}
    \label{fig:model-selection-test2-ref-yc}
  }%
  \caption{Synthetic reference fine and coarse fields, and observation locations, for Test 2}
  \label{fig:model-selection-test2-ref}
\end{figure}

\begin{figure}[tbhp]
  \centering
  \subfloat[Fine scale]{%
    \includegraphics[width=0.75\textwidth]{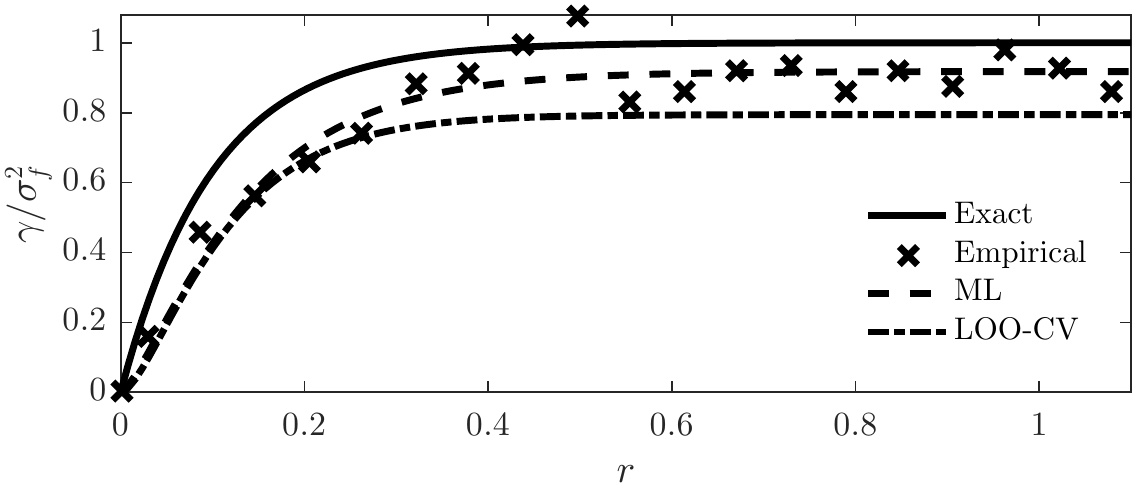}
    \label{fig:model-selection-test2-vgram-f}
  }\\
  \subfloat[Coarse scale]{%
    \includegraphics[width=0.75\textwidth]{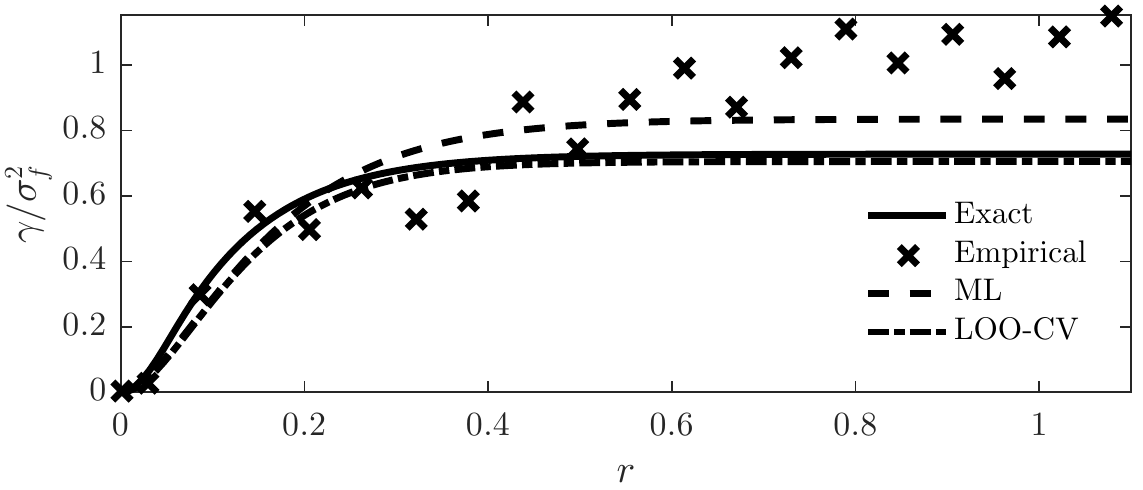}
    \label{fig:model-selection-test2-vgram-c}
  }\\
  \subfloat[Coarse-fine]{%
    \includegraphics[width=0.75\textwidth]{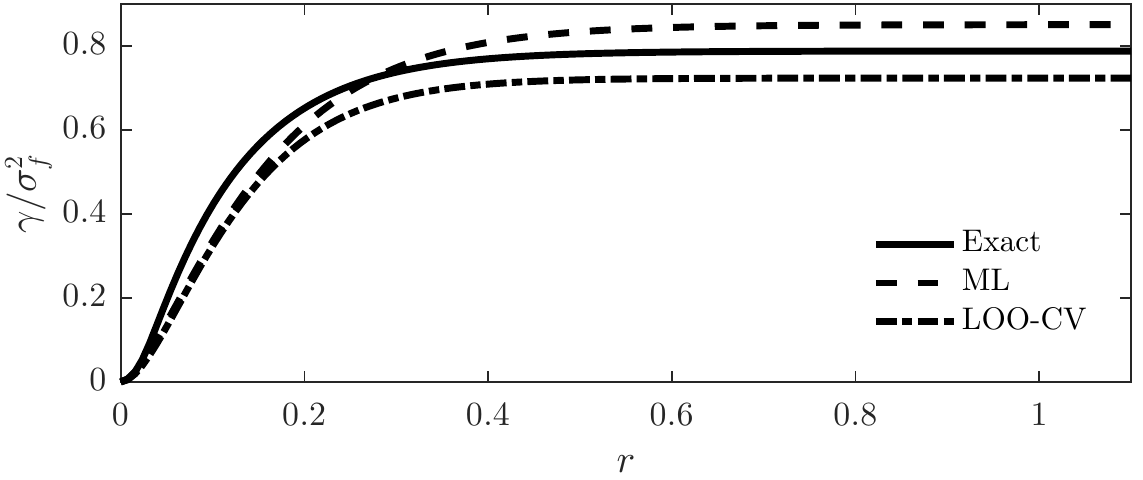}
    \label{fig:model-selection-test2-vgram-cf}
  }%
  \caption{Fine scale and coarse scale variograms, and coarse-fine cross-variograms for Test 2.
    True and estimated variograms indicated in continuous and dashed lines, respectively.
    Empirical variograms indicated with crosses.
  }
  \label{fig:model-selection-test2-vgram}
\end{figure}

\begin{figure}[tbhp]
  \centering
  \subfloat[Conditional mean $\hat{\mu}^f$]{%
    \includegraphics[width=0.436\textwidth]{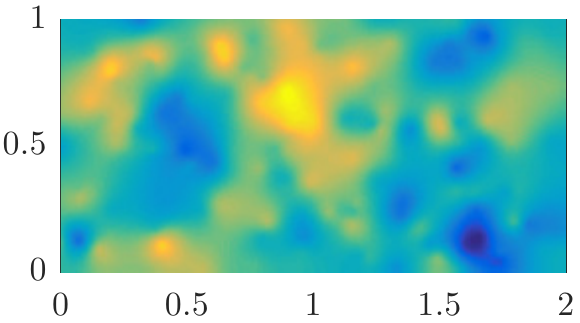}
    \label{fig:model-selection-test2-pred-mean-fine}
  }%
  \subfloat[Conditional mean $\hat{\mu}^c$]{%
    \includegraphics[width=0.544\textwidth]{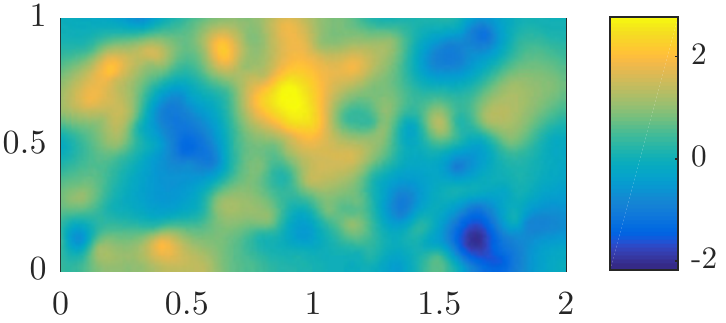}
    \label{fig:model-selection-test2-pred-var-fine}
  }\\
  \subfloat[Conditional variance $(\hat{\sigma}^f)^2$]{%
    \includegraphics[width=0.433\textwidth]{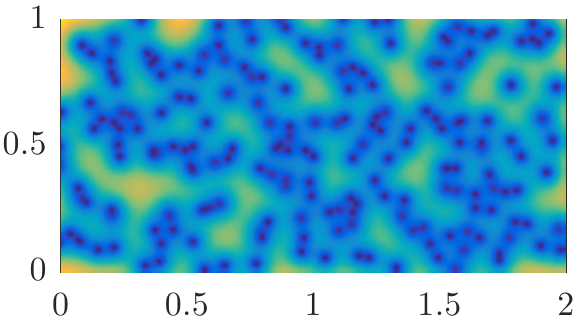}
    \label{fig:model-selection-test2-pred-mean-coarse}
  }%
  \subfloat[Conditional variance $(\hat{\sigma}^c)^2$]{%
    \includegraphics[width=0.547\textwidth]{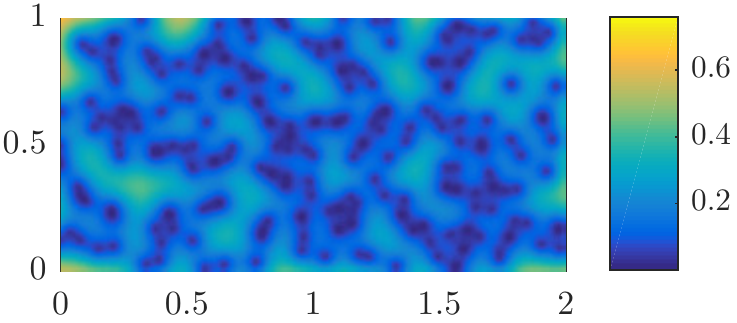}
    \label{fig:model-selection-test2-pred-var-coarse}
  }%
  \caption{Mean and variance of the fine and coarse scale fields, conditioned on multiscale data, for Test 2 %
   }
  \label{fig:model-selection-test2-pred-meanvar}
\end{figure}

For Test 2 we have less differentiation between scales compared to Test 1.
\cref{fig:model-selection-test2-ref} shows both the reference fine and coarse scale fields.
\cref{fig:model-selection-test2-vgram} shows the estimated, true, and empirical variograms.
Additionally, \cref{fig:model-selection-test2-pred-meanvar} shows the conditional mean and variance.
Due to the lesser degree of separation between scales compared to Test 1, model selection is more challenging.
Nevertheless, as for Test 1, we find that both ML and LOO-CV provide estimates for the full bivariate Mat\'{e}rn model that accurately resolve the structure of the multiscale reference model.
We see that both ML and LOO-CV overestimate the shape parameter of the fine scale field, but correctly identify that the coarse field is smoother and has longer correlation length.
As the degree of separation between scales is lesser, we can see in~\cref{fig:model-selection-test1-pred-var-fine} and~\cref{fig:model-selection-test1-pred-var-coarse} that measurements on the fine scale more sharply reduce the conditional variance at the coarse scale, and that measurements on the coarse scale more sharply reduce the conditional variance at the fine scale.

We now illustrate how the estimated hyperparameters of the multivariate Mat\'{e}rn model vary for randomly chosen reference fields and observation locations.
For this experiment we generate $500$ draws of normalized fine scale log-conductivity fields from the same GP and compute for each one the corresponding coarse scale field via block averaging with given support $\eta_c$.
For each pair of fields we obtain hyperparameter estimates using both the ML and LOO-CV approaches from randomly chosen observations.
As in Tests 1 and 2, observation locations are randomly taken from the set of discretization centroids.
The reference field parameters are set as $\lambda_f = 0.05$, $\nu_f = 0.5$, $\sigma_f = 1.0$, $\eta_c = 8$, and $\sigma_{\epsilon} = \num{5e-2}$.
The estimation domain is $\Omega = [0, 1] L \times [0, 1] L$, and fields are computed on a $64 \times 64$ square grid.
$150$ and $50$ observations are taken at the fine and coarse scales, respectively.

\begin{figure}[tbhp]
  \centering%
  \includegraphics[width=0.98\textwidth]{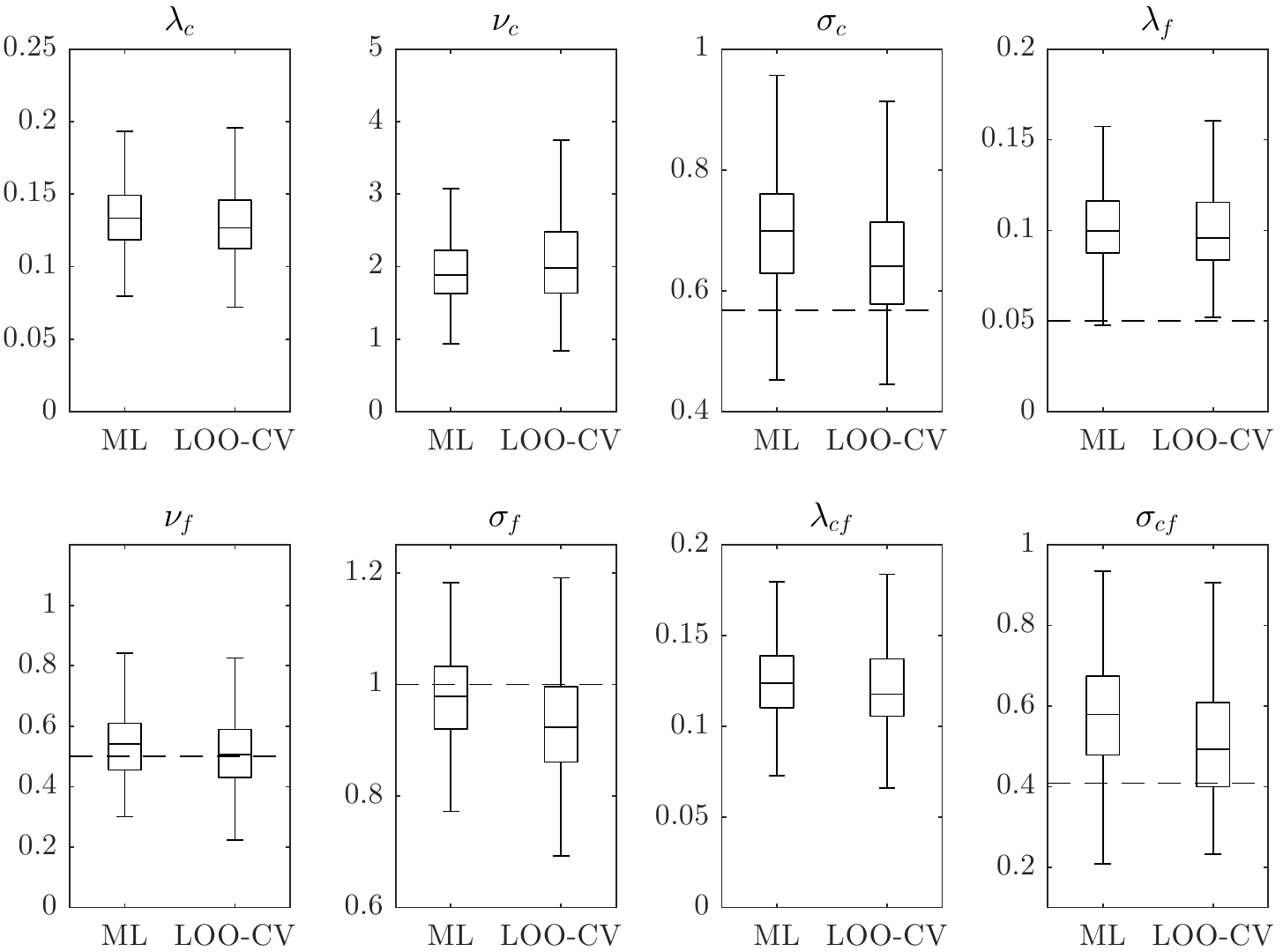}
  \caption{Hyperparameters of the full bivariate model estimated from synthetic multiscale data. with $\lambda_f = 0.05$, $\nu_f = 0.5$, $\sigma_f = 1.0$, and $\lambda_f / \eta_c = 0.4$, estimated from $500$ sets of randomly generated synthetic multiscale data.
    Boxes indicate interquartile range and whiskers extend to the most extreme value that is not 1.5 times the interquartile range from the box.
    Solid lines indicate the median, and dashed lines indicate true values.}
  \label{fig:model-selection-test3}
\end{figure}

Results for Test 3 are summarized in~\cref{fig:model-selection-test3}.
These results, as those of Tests 1 and 2, indicate that the proposed approach is capable of identifying the different degrees of smoothness of each scale.
We also see that the true parameters fall either inside or close the interquartile range of estimated hyperparameters with the exception of $\lambda_f$.
This indicates that more fine scale observations are generally required to accurately estimate the fine scale correlation length.
We finally remark that~\cref{fig:model-selection-test3} indicates that both ML and LOO-CV result in similar estimate variability.
Nevertheless, we recommend the use of LOO-CV estimation as it was more robust with respect to initial guess of hyperparameters.

To close this section, for Tests 1 and 2, we compare the fine and coarse scale fields estimated from multiscale data and estimated from the corresponding single scale datasets.
For this purpose we compute the mean square error (MSE) of the predicted log-conductivity, defined for scale $a = \{ c, f \}$ as
\begin{linenomath*}
  \begin{equation}
  \label{eq:MSE}
  \operatorname{MSE}^a(\mathcal{D}) = \frac{1}{L^2} \int \mathbb{E} \left [ \left ( y^a_{\mathrm{ref}}(\mathbf{x}) - y^a(\mathbf{x}) \mid \mathcal{D} \right )^2 \right ] \, \mathrm{d} \mathbf{x}
  \end{equation}
\end{linenomath*}
The results are summarized in \cref{tbl:model-selection-MSE}.
It can be seen that, for both the fine and coarse scale fields and for both tests 1 and 2, going from single scale data to multiscale data results in a reduction in MSE.
This reduction is more pronounced for the coarse scale field ($24\%$ for test 1 and $33\%$ for test 2).
In fact, we expect fine scale measurements to inform the coarse field estimate more than vice versa due to the indeterminacy of the inverse of the coarsening relation.
Similarly, reduction in MSE is more pronounced for Test 2 than for Test 1, which is to be expected as we observe less scale separation between fine and coarse fields for Test 2.
\begin{table}[tbhp]
  \footnotesize
  \caption{%
    MSE of predicted log-conductivity with respect to the reference for the tests of \cref{tbl:model-selection-tests}, computed using~\cref{eq:MSE}
  }
  \begin{center}\footnotesize
    \begin{tabular}{*{9}{c}}
      \toprule
      & \multicolumn{4}{c}{Single scale data} & \multicolumn{4}{c}{Multiscale data}\\
      & \multicolumn{2}{c}{Fine} & \multicolumn{2}{c}{Coarse} & \multicolumn{2}{c}{Fine}  & \multicolumn{2}{c}{Coarse} \\
      \cmidrule(lr){2-3}%
      \cmidrule(lr){4-5}%
      \cmidrule(lr){6-7}%
      \cmidrule(lr){8-9}%
      Test & ML & LOO-CV & ML & LOO-CV & ML & LOO-CV & ML & LOO-CV \\
      \midrule
      1 & 1.150 & 1.180 & 0.812 & 0.859 & 1.030 & 1.050 & 0.472 & 0.492 \\
      2 & 0.596 & 0.590 & 0.643 & 0.639 & 0.495 & 0.489 & 0.284 & 0.280 \\
      \bottomrule
    \end{tabular}
  \end{center}
  \label{tbl:model-selection-MSE}
\end{table}

\subsection{Uncertainty Propagation}
\label{sec:up}

We present an application of multivariate GP regression to uncertainty propagation in physical systems.
We consider Darcy flow in heterogeneous porous media, where we are interested in predicting spatial hydraulic pressure ($h$) profiles from multiscale observations of $h$ and of the hydraulic conductivity $K$, and estimating the uncertainty in these predictions.
We assume hydraulic head observations have a fine-scale support volume corresponding to piezometric observations, whereas hydraulic conductivity observations have fine-scale and coarse-scale support volumes corresponding to both laboratory experiments performed on field samples and cross-pumping tests.

In order to employ the Darcy flow model to propagate uncertainty in the spatial hydraulic conductivity distribution, both the hydraulic pressure and conductivity fields must be defined with the same support volume.
As $h$ observations are fine-scale, we require constructing a probabilistic model for the fine-scale hydraulic conductivity field $K^f$ from multiscale observations of $K$.
We assume there's a data set $\mathcal{D}_s$ of multiscale observations of the log hydraulic conductivity $Y = \log K$.
In particular we assume the data set of Test 1 in~\cref{sec:numerical-selection}.

The Darcy flow problem reads
\begin{linenomath*}
  \begin{gather}
    \label{eq:darcy-eq}
    \nabla \cdot \left [ K^f(\mathbf{x}) \nabla h^f(\mathbf{x}) \right ] = 0, \quad \mathbf{x} \equiv (x_1, x_2)^{\top} \in \Omega (\equiv [0 \times 2 L ] \times [0, L]), \\
    \label{eq:darcy-eq-bcd}
    h^f(0, x_2) = h^f_{\mathrm{L}}, \quad h^f(2 L , x_2) = h^f_{\mathrm{R}}, \\
    \label{eq:darcy-eq-bcn}
    \frac{\partial}{\partial x_2} h^f(x_1, 0) = \frac{\partial}{\partial x_2} h^f(x_2, L) = 0,
  \end{gather}
\end{linenomath*}
where $h^f(\mathbf{x})$ is the fine-scale hydraulic pressure field.
The field $h^f(\mathbf{x})$ is discretized into the vector $h^f = (h^f(\mathbf{x}_1), \dots, h^f(\mathbf{x}_{N_h}))^{\top}$, where $\{ \mathbf{x}_i \}^{N_h}_{i = 1}$ is a set of discrete nodes.
Our goal is to estimate $h^f(\mathbf{x})$ given discrete observations of $h^f$ and the multiscale data set $\mathcal{D}_s$ of observations of $Y$.
We assume that observations $h^f_s$ of the discrete hydraulic head field are taken as $h^f_s = H h^f + \epsilon_h$, where $H \in \mathbb{R}^{N_{hs} \times N_h}$ is a sampling matrix with $N_{hs} \ll N_h$, and $\epsilon_h \sim \mathcal{N}(0, \sigma^2_{eh} I)$ is the normally distributed observation error.
For this experiment, we take $N_{hs} = 20$ head observations at observation locations randomly chosen from the set of discretization centroids.
Furthermore, we assume an observation error standard deviation $\sigma_{\epsilon h} = \num{5e-2} (h^f_{\mathrm{L}} - h^f_{\mathrm{R}})$.

\begin{figure}[tbhp]
  \centering
  \subfloat[]{%
    \label{fig:up-darcy-unc}
    \includegraphics[width=0.45\textwidth]{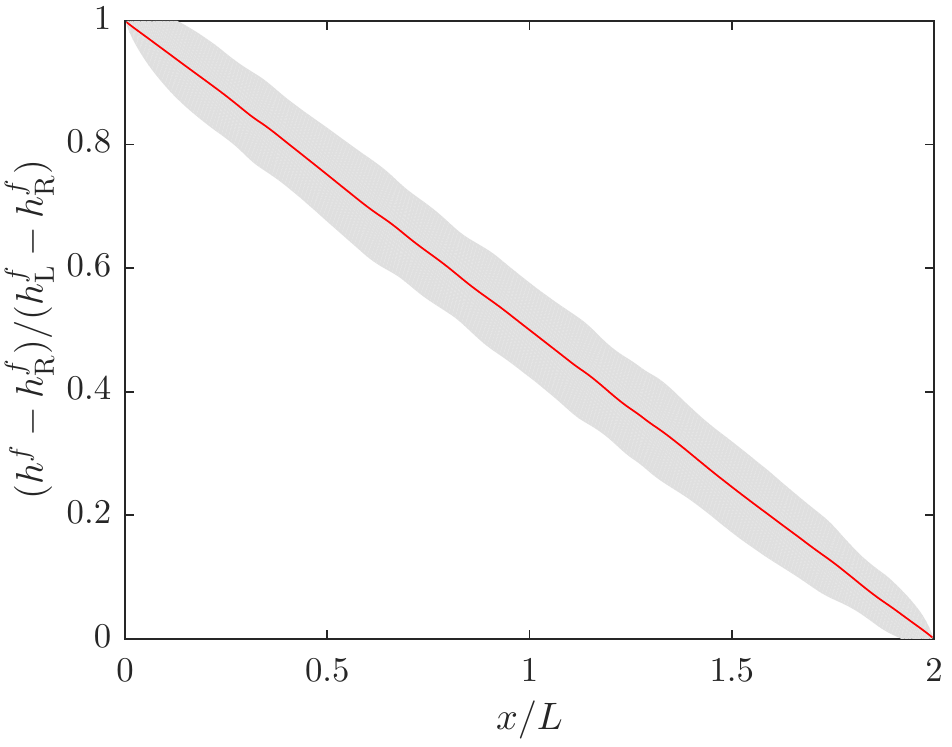}
  }
  \subfloat[]{%
    \label{fig:up-darcy-cond}
    \includegraphics[width=0.45\textwidth]{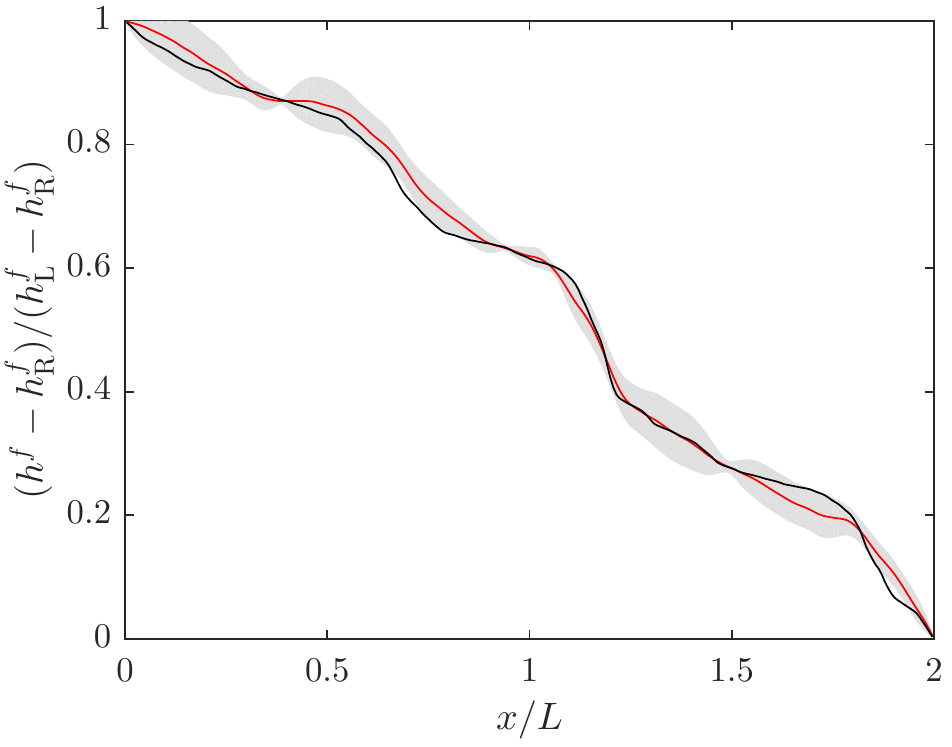}
  }
  \caption{Estimated normalized profile $h^f(x_1, x_2 = 0.5 L)$ computed via minimum mean square estimation.
    \protect\subref{fig:up-darcy-unc} Mean and 95\% confidence interval of Gaussian prior computed from \num{1e3} realizations of the fine-scale log-conductivity field of Test 1 conditioned on multiscale data.
  \protect\subref{fig:up-darcy-cond} Predictive man (red) and 95\% confidence interval, computed using \num{20} head observations, and reference (black)}
  \label{fig:up-darcy}
\end{figure}

We model the discrete log hydraulic conductivity $y$ as a GP.
Given the uncertainty in $y$ and the observations $h^f_s$, the minimum mean square error estimator of the hydraulic head field, $\hat{h}$ (predictive mean), and its covariance $\hat{C}_h$ (predictive covariance), are given by
\begin{linenomath*}
  \begin{align}
    \label{eq:mmse-mean}
    \hat{h} &= \bar{h} + C_h H^{\top} \left ( H C_h H^{\top} + \sigma^2_{\epsilon h} I \right )^{-1} (h^f_s - H \bar{h}), \\
    \label{eq:mmse-covar}
    \hat{C}_h &= C_h - C_h H^{\top} \left ( H C_h H^{\top} + \sigma^2_{\epsilon h} I \right )^{-1} H C_h,
  \end{align}
\end{linenomath*}
where $\bar{h}$ and $C_h$ are the prior mean and covariance of $h^f$ given the uncertainty in $Y$.
We estimate $\bar{h}$ and $C_h$ from \num{1e3} Monte Carlo (MC) simulations of the GP $y^f \mid \mathcal{D}_s$ and the problem~\cref{eq:darcy-eq,eq:darcy-eq-bcd,eq:darcy-eq-bcn}.
Results for the profile $x_2 = 0.5 L$ are presented in \cref{fig:up-darcy} and \cref{tbl:up-var}.

\cref{fig:up-darcy-unc} shows the mean and variance along the profile $x_2 = 0.5 L$.
It can be seen that the variance of $h^f$ due to uncertainty in $Y$ is narrow because of conditioning $y^f$ on multiscale data.
\cref{tbl:up-var} presents the $L^2$-norm of the prior variance along the profile $x_2 = 0.5 L$ obtained when $y^f$ is conditioned on fine-scale data only and on multiscale data.
Consistent with our observations in~\cref{sec:numerical-selection}, conditioning on multiscale data results in a reduction in uncertainty ($\sim 30\%$), this time not only for the parameters but for the state itself.

Employing the MC estimate of $\bar{h}$ and $C_h$, we apply~\cref{eq:mmse-mean,eq:mmse-covar} to obtain the predictive mean and covariance $\hat{h}$ and $\hat{C}_h$ conditioned on $h^f$ observations.
\cref{fig:up-darcy-cond} shows the mean and variance of the estimated hydraulic head along the profile $x_2 = 0.5 L$, together with the reference profile.
Good agreement of the predictive mean $\hat{h}$ with the reference profile is observed, with the reference profile falling inside the $95\%$ confidence interval indicated by the predictive variance.
\cref{tbl:up-var} shows the $L^2$-norm of the conditioned variance when the GP model for $y^f$ is conditioned on fine-scale data and on multiscale data.
It is again seen that by conditioning on multiscale data the uncertainty in the state is reduced.

\begin{table}[tbhp]
  \footnotesize
  \caption{%
    $L^2$-norm of the variance of normalized profile $h^f(x_1, x_2 = 0.5 L)$ prior and after conditioning on $h^f$ observations.
    GP model for $y^f$ trained first on fine-scale data only, and then on multiscale data.
  }
  \begin{center}\footnotesize
    \begin{tabular}{*{3}{c}}
      \toprule
      $h^f(x_1, x_2 = 0.5 L)$ variance & Fine-scale data only & Multiscale data \\
      \midrule
      Prior & \num{1.300e-5} & \num{9.488e-6} \\
      Conditioned & \num{2.210e-6} & \num{1.840e-6} \\
      \bottomrule
    \end{tabular}
  \end{center}
  \label{tbl:up-var}
\end{table}

\section{Conclusions and Future Work}
\label{sec:conclusions}

We have presented a multiscale simple kriging approach for parameter field reconstruction that incorporates measurements with two different support volumes (fine and coarse scales).
Our results show that treating each support scale as components of a bivariate Gaussian process with full bivariate Mat\'{e}rn covariance function is a viable strategy for modeling multiscale observations.
We find that the full bivariate Mat\'{e}rn model with hyperparameters estimated from data accurately captures the stationary features of reference fields for scenarios exhibiting moderate to large separation between fine and coarse scales.
In particular, we show that the model selection approaches employed in this work, marginal likelihood and leave-one-out cross-validation, both qualitatively identify the difference in smoothness and correlation lengths between scales.
Nevertheless, we recommend the use of LOO-CV estimation as it was more robust with respect to initial guess of hyperparameters.

Furthermore, we show that conditioning on multiscale data can be used to reduce uncertainty in stochastic modeling of physical systems with heterogeneous model parameters sampled over multiple scales.

Future work will extend the proposed approach to more than two spatial scales by employing valid multivariate Mat\'{e}rn models~\citep{apanasovich-2012-valid}.

\appendix

\section{Nystr\"{o}m Method for Simulating GPs}
\label{sec:nystrom}

In this section we present the Nystr\"{o}m method employed in this work for simulating GPs.
Nystr\"{o}m methods are used to approximate the factorization of large covariance matrices in terms of the factorization of a smaller covariance matrices~\citep{halko-2011-finding}.

Let $y(\mathbf{x}) \sim \mathcal{GP}(0, C(\mathbf{x}, \mathbf{x}'))$.
Our goal is to simulate realizations of $y$, $y^{(k)}_{*}$ at the $N$ nodes $X_{*}$.
Simulating $y$ via factorization (e.g. Cholesky, eigendecomposition, etc.) is computationally prohibitive for large $N$ as factorizations require $O(N^3)$ operations and $O(N^2)$ storage.
Therefore, we aim to compute a low-rank approximation to the covariance matrix of the form $C(X_{*}, X_{*}) \approx A B^{-1} A^{\top}$, where $B$ is a rank-$M$ spd matrix with $M \ll N$.

By Mercer's theorem, the covariance kernel $C(\mathbf{x}, \mathbf{x}') $ allows the representation in terms of its eigenpairs $\{ \lambda_k, \psi_k \}^{\infty}_{k = 1}$,
\begin{linenomath*}
  \begin{equation}
  \label{eq:mercer}
  C(\mathbf{x}, \mathbf{x}') = \sum^{\infty}_{k = 1} \lambda_k \psi_k(\mathbf{x}) \psi_k(\mathbf{x}'),
  \end{equation}
\end{linenomath*}
where the eigenpairs satisfy the relation
\begin{linenomath*}
  \begin{equation}
  \label{eq:mercer-linop}
  \lambda_i \psi_i(\mathbf{x}) = \int_{\Omega} C(\mathbf{x}, \mathbf{x}') \psi(\mathbf{x}') \, \mathrm{d} \mathbf{x}'
  \end{equation}
\end{linenomath*}
and the orthogonality relation $\int_{\Omega} \psi_i(\mathbf{x}) \psi_j(\mathbf{x}) \, \mathrm{d} \mathbf{x} = \delta_{ij}$.
We approximate the integral in~\cref{eq:mercer-linop} using an $M$-node quadrature rule over $\Omega$ with nodes $\tilde{X}_k = ( \tilde{\mathbf{x}}_1, \dots, \tilde{\mathbf{x}}_M )$ and uniform weight $w$ (e.g., midpoint rule, MC sampling, etc.), obtaining
\begin{linenomath*}
  \begin{equation}
  \label{eq:mercer-nystrom}
  \lambda_i \psi_i(\mathbf{x}) \approx w \sum^M_{k = 1} C(\mathbf{x}, \tilde{\mathbf{x}}_k) \psi_i(\tilde{\mathbf{x}}_k).
  \end{equation}
\end{linenomath*}
Furthermore, let $\tilde{C} \equiv C(\tilde{X}, \tilde{X})$ denote the covariance matrix for the quadrature nodes, with eigendecomposition $\tilde{C} = \tilde{U} \tilde{L} \tilde{U}^{\top}$, $\tilde{\Lambda}= \operatorname{diag}(\tilde{\lambda}_i, \dots, \tilde{\lambda}_M)$.
Approximating $\psi_i(\tilde{\mathbf{x}}_k)$ and $\lambda_i$ in terms of $\tilde{U}$ and $\tilde{\Lambda}$, we obtain
\begin{linenomath*}
  \begin{equation}
  \label{eq:eigenpairs-nystrom}
  \lambda_k \approx w \tilde{\lambda}_k, \quad \psi_i(\mathbf{x}) \approx \frac{1}{w^{1/2} \tilde{\lambda}_i} \sum^M_{k = 1}C(\mathbf{x}, \tilde{\mathbf{x}}_k) \tilde{U}_{ki}, \quad k = 1, \dots, M.
  \end{equation}
\end{linenomath*}
The relations in~\cref{eq:eigenpairs-nystrom} can be employed to construct an approximate K-L expansion, which then can be used to simulate the field.
In this work we follow an alternative procedure, employed in~\cref{sec:up} for simulation and in~\cref{sec:numerical-selection} to compute conditional mean and variances.
Substituting~\cref{eq:eigenpairs-nystrom} into~\cref{eq:mercer}, we obtain the following rank-$M$ approximation to the covariance matrix $C(X_{*}, X_{*})$,
\begin{linenomath*}
  \begin{equation}
  \label{eq:covar-nystrom}
  C(Y, Z) \approx C(Y, \tilde{X}) \left ( \tilde{U} \tilde{\Lambda}^{-1} \tilde{U}^{\top} \right ) C(\tilde{X}, Z) = C(Y, \tilde{X}) \tilde{C}^{-1} C(\tilde{X}, Z),
  \end{equation}
\end{linenomath*}
where $Y$ and $Z$ are two arbitrary sets of nodes.

For prediction, we employ \cref{eq:covar-nystrom} to compute $C^c_p$ (and $C^f_p)$ in~\cref{eq:msk-cond-p-coarse}.
For simulation, we note that the covariance matrix $\tilde{C}$ admits the Cholesky factorization $\tilde{C} = \tilde{L} \tilde{L}^{\top}$.
Substituting this factorization into~\cref{eq:covar-nystrom} for $Y = Z = X_{*}$ we obtain the rectangular factorization
\begin{linenomath*}
  \begin{equation}
  \label{eq:covar-nystrom-fac}
  C(X_{*}, X_{*}) = L L^{\top}, \quad L^{\top} = \tilde{L}^{-1} C(\tilde{X}, X_{*}).
  \end{equation}
\end{linenomath*}
The realizations are simulated as $y^{(k)}_{*} = L \xi^{(k)}$, where the $\xi^{(k)}$ are realizations of $\xi \sim \mathcal{N}(0, I_M)$.
This procedure requires $O(N M^2)$ operations and $O(MN)$ storage.

We employed the centroids of a $128 \times 64$ ($M = \num{8192}$) uniform discretization of $\Omega$ as quadrature nodes for computing conditional mean and variance in \cref{sec:numerical-selection} and for MC simulations presented in~\cref{sec:up}.

\acknowledgments%
All code employed to generate all data and results presented in this manuscript can be found at \url{https://doi.org/10.5281/zenodo.1205644}.
This research at Pacific Northwest National Laboratory (PNNL) was supported by the U.S. Department of Energy (DOE) Office of Advanced Scientific Computing Research (ASCR) as part of the project ``Uncertainty quantification for complex systems described by Stochastic Partial Differential Equations.''
PNNL is operated by Battelle for the DOE under Contract DE-AC05-76RL01830.

\end{document}